\documentclass[11pt]{article}
\usepackage[utf8]{inputenc}
\usepackage[T1]{fontenc}
\usepackage{amsmath,amssymb}
\usepackage{booktabs}
\usepackage[margin=1in]{geometry}
\usepackage{authblk}
\usepackage[protrusion=true,expansion=false]{microtype}
\usepackage[hidelinks]{hyperref}
\usepackage{caption}
\title{\bfseries Two-stage imputation of longitudinal anthropometric data with cross-reference harmonisation: a simulation study}
\author[1]{Flavia Alves\thanks{Corresponding author. Department of Public Health, Policy and Systems, Whelan Building, University of Liverpool, Liverpool L69 3GB, United Kingdom. ORCID: 0000-0001-5826-2976.}}
\affil[1]{Department of Public Health, Policy and Systems, University of Liverpool, Liverpool, United Kingdom}
\date{}

\begin{document}
\maketitle

\begin{abstract}
\noindent\textbf{Objective.} Longitudinal datasets frequently contain missing weight and height measurements, and studies that combine data sources may index measurements against different growth reference standards (e.g., the WHO reference and CDC charts). We describe and evaluate a reproducible two-stage method that imputes missing anthropometry while making the choice of reference standard an explicit parameter.

\noindent\textbf{Methods.} Stage~1 applies within-subject linear interpolation across visit dates (interior gaps only, no extrapolation). Stage~2 imputes remaining values from an age- and sex-specific growth reference using the LMS method by estimating each subject's centile, carrying it forward and backwards within the subject, defaulting to the 50th centile when a subject is never measured, and reading the expected value off the reference at the visit age. Different references can be supplied per data source so that the standard applied is recorded and auditable. We assessed recovery accuracy by masking and re-imputing a random 20\% of observed values. All evaluations used computer-generated synthetic data.

\noindent\textbf{Results.} On synthetic data ($n=60$ subjects, 288 visits, 30\% missing), the method resolved missingness to 100\% completeness. Masked-value recovery gave a mean absolute error of 1.78~kg for weight (3.5\% mean absolute percentage error) and 2.84~cm for height (2.0\%), with negligible bias. Values recovered by within-subject interpolation were more accurate than those recovered from the growth reference, as expected, supporting the two-stage ordering.

\noindent\textbf{Conclusion.} The method offers a simple, dependency-free, and auditable approach to anthropometric imputation, with explicit handling of differing reference standards and per-value provenance. Application to empirical data and propagation of imputation uncertainty into downstream models are the necessary next steps before use in substantive analyses.
\end{abstract}


\newpage

\section*{Significance and Innovations}
\begin{itemize}
\item Combining data indexed against \textbf{different growth reference standards} (e.g.\ WHO vs CDC) introduces a systematic, often unstated, difference; this method makes the reference standard an \textbf{explicit, auditable parameter} rather than a hidden assumption.
\item A \textbf{two-stage design} prioritises within-subject information (linear interpolation) before falling back to population growth references and records the provenance of every imputed value so analysts can stratify or exclude by source.
\item The method is implemented in \textbf{base statistical software with no package dependencies} and is fully reproducible on synthetic data, lowering the barrier to reuse.
\item The report is transparent about the method's key assumption --- \textbf{centile stability over time} --- which may not hold in individuals undergoing rapid change, and shows how provenance labelling helps mitigate it.
\end{itemize}

\section{Introduction}
Anthropometric measurements --- weight and height --- are fundamental variables in health research, underpinning dosing, risk modelling, and growth monitoring. In real-world longitudinal datasets, these measurements are frequently missing at individual time points due to non-standardised data collection, missed assessments, or selective recording~\cite{sterne2009,little2019}. Na\"ive handling --- complete-case analysis or last-observation-carried-forward --- can bias estimates and discard information, particularly when missingness is related to the state being measured~\cite{white2011}.

Imputing anthropometric measurements in growing individuals is complicated because plausible values depend jointly on age and sex; values cannot be imputed from a single fixed distribution. Growth references address this by expressing a measurement as an age- and sex-specific centile or standard-deviation score via the LMS method~\cite{cole1992}, in which the skewness ($L$), median ($M$), and coefficient of variation ($S$) vary smoothly with age. Two widely accepted standards --- the WHO growth reference~\cite{deonis2007} and the CDC growth charts~\cite{cdc2000} --- differ in their reference populations and therefore assign different centiles to the same measurement. When data indexed against different standards are combined, the choice of reference becomes a substantive decision that is often left implicit.

We describe a two-stage imputation method that (i) exploits within-subject longitudinal structure before resorting to population references, (ii) makes the reference standard an explicit, source-specific parameter so that harmonisation is auditable, and (iii) labels the provenance of every imputed value. We evaluate recovery accuracy using a mask-and-recover design on synthetic data. This is a methodological report: all results are generated on computer-simulated data and are intended to demonstrate the method's behaviour and reproducibility.

\section{Methods}
\subsection*{Synthetic data}
We generated a synthetic dataset reproducing the structure of a longitudinal study: repeated visits per subject, two data sources with distinct reference standards, an imbalanced sex distribution, realistic missingness, and a small number of gross outliers. For each synthetic subject, a latent weight centile and height centile were drawn and used with an assigned reference to generate measurements at each visit age, with multiplicative noise. The generation procedure is described here in sufficient detail to reproduce equivalent datasets and uses standard pseudo-random sampling.

\subsection*{Stage 1: within-subject linear interpolation}
For each subject, observed values of a measure were ordered by visit date and missing interior values were filled by linear interpolation between adjacent observed measurements. No extrapolation was performed: values before the first or after the last observation were left for Stage~2. Interpolation therefore uses only a subject's own measurements and is preferred wherever two surrounding observations exist.

\subsection*{Stage 2: growth-reference (LMS) centile imputation}
Values still missing after Stage~1 were imputed from an age- and sex-specific growth reference. For each observed measurement we computed the LMS standard-deviation score and corresponding centile, where for $L \neq 0$ the score is
\[
z = \frac{(\text{value}/M)^{L} - 1}{L\,S},
\]
and the inverse transform
\[
\text{value} = M\,(1 + L\,S\,z)^{1/L}
\]
returns the expected measurement at a target centile~\cite{cole1992}. Each subject's observed centile was carried forward and backward across visits; subjects with no observed value for a measure were assigned the 50th centile by default. The expected measurement was then read off the reference at the visit's age and sex. This assumes approximate centile stability within a subject.

\subsection*{Cross-reference handling}
To handle data indexed to different standards, the reference table is supplied per source: in our demonstration, one source was indexed to a WHO-style reference and the other to a CDC-style reference. The choice is an explicit argument to the procedure, so the standard applied to each subject is recorded and auditable.

\subsection*{Outlier handling and provenance}
Before imputation, implausible values were flagged using an age-banded robust $z$-score (median and median absolute deviation within two-year age bands) and set to missing when the absolute robust $z$-score exceeded 3; this threshold is configurable, and a fixed biological-limits rule can be substituted. Each final value was labelled by source --- observed, interpolated, or growth-reference --- enabling downstream stratification or sensitivity analysis.

\subsection*{Evaluation}
We assessed recovery accuracy by randomly masking 20\% of observed values, imputing them with the full pipeline, and comparing the imputed values with the held-out true values using mean absolute error (MAE), root mean squared error (RMSE), mean error (bias), and mean absolute percentage error (MAPE), overall and by imputation source. Analyses used base R~4.3~\cite{rcore}.

\section{Results}
The synthetic dataset comprised 60 subjects and 288 visits, with 30\% of weight and height values missing and 4 gross outliers introduced. After imputation, completeness reached 100\%. Of the weight values, 198 were observed, 36 interpolated, and 54 imputed from the growth reference charts (Table~\ref{tab:prov}).

In the mask-and-recover evaluation (40 held-out values per measure; Table~\ref{tab:acc}), weight was recovered with an MAE of 1.78~kg (RMSE 2.18~kg; MAPE 3.5\%) and height with an MAE of 2.84~cm (RMSE 3.70~cm; MAPE 2.0\%). Bias was negligible for both (weight: $-0.07$~kg; height: $+0.17$~cm). As expected, values recovered by within-subject interpolation were more accurate than those recovered from the growth reference (weight: 1.47 vs 2.06~kg; height: 2.60 vs 2.99~cm), supporting the two-stage ordering.

\begin{table}[htbp]
\centering
\caption{Imputation provenance for weight in the synthetic demonstration dataset.}
\label{tab:prov}
\begin{tabular}{lrl}
\toprule
Source & $n$ values & Interpretation \\
\midrule
Observed & 198 & Recorded during the visit \\
Interpolated & 36 & Stage 1, within-subject linear interpolation \\
Growth reference & 54 & Stage 2, LMS centile imputation \\
Total & 288 & 100\% complete after imputation \\
\bottomrule
\end{tabular}
\end{table}

\begin{table}[htbp]
\centering
\caption{Mask-and-recover accuracy (20\% of observed values held out).}
\label{tab:acc}
\begin{tabular}{lrrrrr}
\toprule
Measure & $n$ masked & MAE & RMSE & Bias & MAPE \\
\midrule
Weight (kg) & 40 & 1.78 & 2.18 & $-0.07$ & 3.5\% \\
Height (cm) & 40 & 2.84 & 3.70 & $+0.17$ & 2.0\% \\
\bottomrule
\end{tabular}

\vspace{5pt}
{\small MAE by imputation source (interpolation / growth reference): weight 1.47 / 2.06~kg; height 2.60 / 2.99~cm.}
\end{table}

\newpage

\section{Discussion}
We have described a two-stage method for imputing longitudinal anthropometry that prioritises within-subject information, falls back to age- and sex-specific growth references via the LMS method, and treats the reference standard as an explicit, source-specific parameter. In a fully synthetic demonstration, the method restored complete data and recovered held-out values with low error and negligible bias, and the two-stage ordering behaved as intended, with interpolation outperforming growth-reference imputation.

The main methodological contribution is making cross-standard handling explicit. When data indexed against WHO and CDC references are combined, the same measurement maps to different centiles; leaving this implicit can introduce a systematic artefact. By requiring that the reference be specified per source and recorded, the method renders this decision auditable and reversible in sensitivity analyses.

Several limitations warrant emphasis. First, Stage~2 assumes approximate centile stability within a subject over time. In individuals undergoing rapid or pathological change in body size, centiles can shift markedly, so growth-reference-imputed values should be interpreted cautiously; the provenance label allows analysts to identify and stratify these values. Second, the method produces single-point predictions. Treating imputed values as if observed in a downstream inferential model understates uncertainty and can bias standard errors; multiple imputation or explicit propagation of prediction uncertainty is the appropriate next step~\cite{white2011}. Third, mixing reference standards, while the point of the harmonisation, remains a modelling choice that should be reported and, ideally, examined under alternative standards. Finally, all results here are from synthetic data; they demonstrate behaviour and reproducibility but do not establish accuracy in any empirical setting, which requires validation against observed measurements in real datasets.

Despite these caveats, the approach offers a simple, dependency-free, and fully reproducible tool for a common problem. By coupling longitudinal and reference-based information with explicit standard selection and per-value provenance, it supports both principled imputation and transparent reporting. Future work should validate recovery accuracy on empirical data, compare alternative references head-to-head, and integrate the imputations into uncertainty-aware downstream models.

\end{document}